\documentclass[11pt]{article}
\usepackage{amsmath, amsthm, amscd, amsfonts, amssymb, graphicx, color}
\usepackage[bookmarksnumbered, colorlinks]{hyperref} \usepackage{float}
\usepackage{lipsum}
\usepackage{afterpage}
\usepackage[labelfont=bf]{caption}
\usepackage[nottoc,notlof,notlot]{tocbibind}
\usepackage{fancyhdr}
\newcommand{\keyword}[1]{\par\noindent \textbf{Keywords:} #1 }

\title{Kinematics and Generalized Raychaudhuri equation  in 
 $f(\mathcal{G})$ gravity in imperfect fluid case}
\author
{Y. Alipour Fakhri\thanks{Corresponding Author:
Faculty of Basic Sciences,
Department of Mathematics, Payame Noor University, Tehran, Iran.  E-mail: y\_ alipour@pnu.ac.ir}
\and
Mojtaba Safdarian\thanks{Department of Complementary Education, Payam Noor University, Tehran, Iran.
E-mail: mojtaba\_safdarian@pnu.ac.ir}
\and S. Zamani Moghaddam \thanks{Department of Physics,
Faculty of Science, University of Kashan, P.O. Box 87317-53153,
Kashan, Iran. E-mail:s.zamanimoghadam@gmail.com}}
\begin{document}
\maketitle
\begin{abstract}
The equation of motion is the important equation for obtain the extra force and Raychaudhuri equation. By considering an explicitly coupling between an arbitrary function of the scalar Gauss-Bonnet, $\mathcal{G}$ and the Lagrangian density of matter, it is shown that an extra force normal to their four-velocities arises. In this paper, we obtain  the extra force  and the generalized Raychaudhuri equation in $F(\mathcal{G})$ modified theory of gravity in an imperfect fluid for the massive particle by divergence of energy-momentum tensor  so we earn extra force an Raychaudhuri equation  in a compared with $f(R)$ modified gravity for perfect fluid this conclusion giving the evolution of the kinematical quantities and describing the relative accelerations of nearby particles . 
\end{abstract}
 \keyword{$f(\mathcal{G})$ modified gravity, extra force ,Raychaudhuri equation, Non-minimal coupling constant, Imperfect fluid.}
\section{Introduction}
The $f(\mathcal{G})$ Gauss-Bonnet modified gravity is one of the modified gravity model that is proposed recently to explain physical phenomena such as late time accelerated expansion of the universe without the need of dark energy 
(see \cite{Puxun Wu, S. Nojiri}). The Raychaudhuri equation is a well known and useful equation in general relativity and cosmology . This equation proof the singularity theorems due to Penrose Hawking and Geroch \cite{N. Dadhich}. The Raychaudhuri’s equation is one of the most important tools in investigating the evolution of such geodesics. The geodesic deviation equation describing the relative accelerations of nearby particles, and the Raychaudhury equation giving the evolution of the kinematical quantities associated with deformations (expansion, shear and rotation) are considered in the framework of modified theories of gravity with an arbitrary curvature matter coupling, by taking into account the effects of the extra force (for review, see 
\cite{N. Dadhich,T. Harko, M. Sharifa}). That generalization of this equation was obtained in several forms in some applications for example for time like and null geodesics and for spinning and hot gravitating fluids and in strings and membranes and in different vectors \cite{S. Kar, R. M. Wald}. The generalized Raychaudhuri equation substituted in absence of an extra force \cite{O. Gron}  and it calculated in $f(R)$ modified gravity with non minimal coupling constant \cite{T. Harko}. We will obtain this generalizing equation by calculating the extra force in $f(\mathcal{G})$ modified gravity with non minimal coupling constant in imperfect fluid in a compared with $f(R)$ modified gravity for perfect fluid. 
\section{Equation of motion in $f(\mathcal{G})$ with non-minimal coupling gravity in imperfect fluid}
We begin with an action for the model of modified 
$f(\mathcal{G})$ Gauss-Bonnet  gravity with non-minimal coupling constant $\lambda$ (see \cite{S. Nojiri})
\begin{align}\label{1}
S=\int\sqrt{-g}\Big(\frac{1}{2}R+f(\mathcal{G})+
(1+\lambda F(\mathcal{G})L)\Big)d^4x,
\end{align}
where $F(\mathcal{G})$ is an arbitrary function of
$\mathcal{G}=R^2-4R_{\alpha\beta}R^{\alpha\beta}+
R_{\alpha\beta\tau\nu}R^{\alpha\beta\tau\nu}$, $\lambda$
is a coupling constant and $L$ is a matter Lagrangian.
Varying the action with respect to the metric $g_{\mu\nu}$ yield the field equation
\begin{align}\label{2}
\nonumber
(1+\lambda F)T_{\mu\nu}=R_{\mu\nu}-\frac{1}{2}Rg_{\mu\nu}-
g_{\mu\nu}f+4[RR_{\mu\nu}+R_\mu^{\alpha\beta\gamma}
R_{\nu\alpha\beta\gamma}-2R_\mu^\alpha R_{\alpha\nu}\\
\nonumber
+2R_{\mu\alpha\beta\nu}R^{\alpha\beta}-2G_{\mu\nu}\nabla^2
-R\nabla_mu\nabla_\nu-2g_{\mu\nu}R_{\alpha\beta}\nabla^\alpha
\nabla^\beta\\
+2R_{\alpha\nu}\nabla^\alpha\nabla_\mu
+2R_{\mu\alpha}\nabla^\alpha\nabla_\nu-2R_{\mu\alpha\beta\nu}
\nabla^\alpha\nabla^\beta(f'+\lambda LF')]
\end{align}
where $G_{\mu\nu}$  is the Einstein tensor, 
$T_{\mu\nu}=\frac{2}{\sqrt{-g}}\frac{\delta(L\sqrt{-g})}
{\delta g^{\mu\nu}}$ is the energy-momentum tensor, $f$,
$f'$, $F$ and $F'$ stand for $f(\mathcal{G})$,  
$\frac{df(\mathcal{G})}{d\mathcal{G}}$, $F(\mathcal{G})$ and 
$\frac{dF(\mathcal{G})}{d\mathcal{G}}$ respectively, and
\begin{align}\label{3}
\nonumber
H_{\mu\nu}=RR_{\mu\nu}+ R^{\alpha\beta\gamma}_\mu
R_{\nu\alpha\beta\gamma}-R_\mu^\alpha R_{\alpha\nu}
+2R_{\mu\alpha\beta\nu}R^{\alpha\beta}-2G_{\mu\nu}\nabla^2-R
\nabla_\nu\nabla_\mu\\
-2g_{\mu\nu}R_{\alpha\beta}\nabla^\alpha\nabla^\beta
+2R_{\alpha\nu}\nabla^\alpha\nabla_\mu
+2R_{\mu\alpha}\nabla^\alpha\nabla_\nu
-2R_{\mu\alpha\beta\nu}\nabla^\alpha\nabla^\beta.
\end{align}
So we have definitely  
\begin{align}\label{02}
(1+\lambda F)T_{\mu\nu}=G_{\mu\nu}-g_{\mu\nu}f+4H_{\mu\nu}
(f'+\lambda LF').
\end{align}
Now taking the divergence of both sides of equation (\ref{2}) yields (for review, see \cite{M. Mohseni})
\begin{align}
\nonumber
\nabla^\nu T_{\mu\nu}=\frac{\lambda F'}{1+\lambda F}(g_{\mu\nu}L-
T_{\mu\nu})\nabla^\nu\mathcal{G}\\
+[8RR_{\mu\nu}+
4R^{\alpha\beta\gamma}_\mu R_{\nu\alpha\beta\gamma}-16R_\mu^\alpha
R_{\alpha\nu}+8R_{\mu\alpha\beta\nu}R^{\alpha\beta}]\nabla^\nu L.
\end{align}
So we have
\begin{align}\label{4}
\nabla^\nu T_{\mu\nu}=\frac{\lambda F'}{1+\lambda F}\Big(
(g_{\mu\nu}L-T_{\mu\nu})\nabla^\nu \mathcal{G}+K_{\mu\nu}
\nabla^\nu L\Big).
\end{align}
Here
$K_{\mu\nu}=8RR_{\mu\nu}+4R^{\alpha\beta\gamma}_\mu
R_{\nu\alpha\beta\gamma}-16R^\alpha_\mu R_{\alpha\nu}+
8R_{\mu\alpha\beta\nu} R^{\alpha\beta}$.

Then we consider the energy momentum tensor of an imperfect  fluid
\cite{C.A.Kolassis}
\begin{align}\label{5}
T_{\mu\nu}=(\rho+p-\xi\theta)u_\mu u_\nu+(p-\xi\theta)g_{\mu\nu} -2n\sigma_{\mu\nu}+u_\mu q_\nu+u_\nu q_\mu,  
\end{align} 
where $\rho$ is the overall energy density and $p$ is the isotropic
pressure respectively. The four velocity $u^\mu$ satisfies the conditions $u_\mu u^\mu=1$ and $u_\mu u^{\mu;\nu}=0$. The heat conduction is described by the heat flux vector $q^\alpha$ defined as follows (for review, see \cite{C.A.Kolassis})
\begin{align}
q_\alpha u^\alpha=0.
\end{align}
Also
$n\geq 0$ is the coefficient of dynamic viscosity and $\xi\geq 0$
is the coefficient of bulk viscosity. The quantities
\begin{align}\label{6}
\theta=u^\alpha_{\ ;\alpha}
\end{align}
\begin{align}\label{7}
\sigma_{\alpha\beta}=u_{(\alpha;\beta)}+\dot{u}_{(\alpha} u_{\beta)}-\frac{1}{3}\theta(g_{\alpha\beta}+u_\alpha u_\beta),
\end{align}
are the expansion and shear velocity of the fluid which, according to (\ref{7}), satisfies the condition $u^\alpha \sigma_{\alpha\beta}=0$ (where the round brackets on the indices denote symmetrization, the semicolon denotes
covariant differentiation and the dot denotes differentiation in the direction of $u^\alpha$).

Now we introduce the projection operator 
$h^{\mu\nu}=g^{\nu\mu}-u^\mu u^\nu$. By contracting (\ref{5}) with the projection operator $h^{\lambda\mu}$ we obtain 
\begin{align}\label{9}
\nonumber
(\rho+p-\xi\theta)g^{\mu\alpha}u_\nu\nabla^\nu u_\mu-
\nabla^\nu(p-\xi\theta)(\delta^\alpha_\nu-u_\nu u^\alpha)-2n
\nabla^\nu\sigma_\nu^\alpha\\
+\nabla^\nu(u_\nu q^\alpha+u^\alpha
q_\nu)-\frac{\lambda F'}{1+\lambda F}\Big[
(L+p)\nabla^\nu\mathcal{G}
(\delta^\alpha_\nu-u_\nu u^\alpha)
-K_{\mu\nu}\nabla^\alpha L\Big]=0.
\end{align}
Finally  contraction with $g^{\alpha\beta}$ gives rise to the equation of motion for a fluid element
\begin{align}\label{10}
\frac{Du^\alpha}{ds}=\frac{du^\alpha}{ds}+\Gamma^\alpha_{\mu\nu}
u^\mu u^\nu=f^\alpha.
\end{align}
where the functions $\Gamma^\alpha_{\mu\nu}$ are the Christoffel symbols. Then we obtain the extra force of $F(\mathcal{G})$ with non-minimal coupling gravity for a massive particle 
\begin{align}\label{11}
\nonumber
f^\alpha=\frac{1}{\rho+p-\xi\theta}\Big[
\nabla_\nu(p-\xi\theta)(g^{\alpha\nu}-u^\alpha u^\nu)+2n
\nabla_\nu\sigma^{\alpha\nu}\\
-\nabla_\nu(u^\nu q^\alpha+u^\alpha q^\nu)
+\frac{\lambda F'}{1+\lambda F}\Big((L+p)\nabla_\nu(g^{\alpha\nu}
-u^\alpha u^\nu)-g^\alpha_\mu K^{\mu\nu}\nabla_\nu L\Big)\Big].
\end{align} 
Equation (\ref{11}) may be compared with the one obtained in 
\cite{O. Bertolami} for $f(R)$ gravity.
As one can immediately verify the extra force $f^\alpha$ is orthogonal to the four velocity $u_\alpha$ of the particle, that is, $f^\alpha u_\alpha=0$ (see \cite{O. Bertolami}).

Notice that massless particles do follow geodesic trajectory  and therefore for then $f^\alpha=0$ \cite{N. Dadhich} that may be compared with equation (\ref{11}) in \cite{O. Bertolami}. 
\section{Generalized Raychaudhuri equation in $f(\mathcal{G})$ gravity in imperfect fluid case}
The Raychaudhuri equation can be written in the presence of an extra force as follows \cite{R. M. Wald, O. Gron}
\begin{align}\label{12}
\dot{\theta}+\frac{1}{3}\theta^2+\sigma^2-\omega^2=\nabla_\mu
f^\mu-R_{\mu\nu}u^\mu u^\nu,
\end{align}
where $\sigma^2=\sigma_{\mu\nu}\sigma^{\mu\nu}$ and 
$\omega^2=\omega_{\mu\nu}\omega^{\mu\nu}$. Here $\sigma_{\mu\nu}$ is the shear tensor, $\omega_{\mu\nu}$ is the vorticity tensor,  $\theta$ is the expansion scaler and $R_{\mu\nu}$ is the Ricci tensor.

By using the expression of  the extra force in $f(\mathcal{G})$ gravity in equation (\ref{11}), we obtain the generalized \emph{ generalized Raychaudhuri equation} in imperfect fluid case in 
$f(\mathcal{G})$ gravity as follows
\begin{align}\label{13}
\nonumber
\dot{\theta}+\frac{1}{3}\theta^2+\sigma^2-\omega^2=
\frac{1}{\rho+p-\xi\theta}
\nabla_\mu\Big[
\nabla_\nu(p-\xi\theta)(g^{\alpha\nu}-u^\alpha u^\nu)
+2n\nabla_\nu\sigma^{\alpha\nu}\\
\nonumber
-\nabla_\nu(u^\nu q^\alpha+u^\alpha q^\nu)
+\frac{\lambda F'}{1+\lambda F}\Big((L+p)\nabla_\nu(g^{\alpha\nu}
-u^\alpha u^\nu)-g^\alpha_\mu K^{\mu\nu}\nabla_\nu L\Big)\Big]\\
-R_{\mu\nu}u^\mu u^\nu.
\end{align} 
 The vorticity  $\omega_{\mu\nu}$ satisfies the equation 
 (see \cite{O. Gron, R. M. Wald})
 \begin{align}
 \frac{d\omega_{\mu\nu}}{ds}=-\frac{2}{3}\theta\omega_{\mu\nu}-2
 \sigma^\alpha_{[\nu} \omega_{\mu]\alpha}+\nabla_{[\mu f_\nu]}.
 \end{align}
 While the dynamics  of the  $\sigma_{\mu\nu}$ is described by the equation
 \begin{align}
\nonumber
\frac{d\sigma_{\mu\nu}}{ds}=\frac{1}{2}h^\alpha_\mu h^\beta_\nu
R_{\alpha\beta}+h^\alpha_\mu h^\beta_\nu \nabla_{[\alpha f_\beta]}
-\frac{2}{3}\theta\sigma_{\mu\nu}-
\omega_{\mu\alpha}\omega^\alpha_\nu\\
-\sigma_{\mu\alpha}\sigma^\alpha_\nu-\frac{1}{3}h_{\mu\nu}\Big(
\omega^2-\sigma^2+\frac{1}{2}h^{\alpha\beta}R_{\alpha\beta}
+\nabla_\alpha f^\alpha\Big)-C_{\mu\alpha\nu\beta}u^\alpha u^\beta,
\end{align}
where $C_{\mu\alpha\nu\beta}$ is the Weyl tensor.
It can be showed easily that the generalized equation in the non-presence of extra force state, clearly coincidences with the standard Raychaudhuri equation in general relativity, that is in the state geodesic trajectories
\begin{align}
\dot{\theta}+\frac{1}{3}\theta^2+\sigma^2-\omega^2+R_{\mu\nu},
u^\mu u^\nu=0
\end{align}
\begin{align}
\frac{Du^\alpha}{ds}=\frac{du^\alpha}{ds}+\Gamma^\alpha_{\mu\nu}
u^\mu u^\nu=0.
\end{align}
Existence of extra force is non geodesic trajectory state and geodesic trajectory is equals to don’t exist an extra force. 
\section{Discussion}
In this work we can showed that in $f(\mathcal{G})$ modified theories of gravity with non-minimal coupling of an arbitrary function of the Gauss-Bonnet curvature to the matter Lagrangian, test particles move a long non-geodesic trajectories due to the extra force originated from the coupling and so extra force are used in the Raychaudhuri equation.  
We started with an action of $f(\mathcal{G})$ Gauss-Bonnet gravity with non-minimal coupling constant in an imperfect fluid. Its variation with respect to the metric lead to a new equation of motion. We then obtained an extra force in an imperfect fluid 
simultaneously. Also we obtain generalized Raychaudhuri equation in $f(\mathcal{G})$ Gauss-Bonnet gravity with non-minimal coupling constant for a  massive particle in imperfect fluid. With the use of the expression of the extra force $f^\mu$, the evolution equations for the shear and vorticity can also be obtained explicitly for modified  $F(G)$ gravity with an arbitrary curvature matter coupling. It can be showed easily that the generalized equation in the non-presence of extra force state, clearly coincidences with the standard Raychaudhuri equation in general relativity in the state of geodesic trajectories that don't have any extra force. Maybe extra force in modified theories of gravity can be predict and candidate theories that describe dark matter too base of experimental observation .
\\*
\textbf{DATA AVAILABILITY:}
The data that support the findings of this study are available from the corresponding author upon reasonable request.

\end{document}